\documentstyle[12pt,aasms4]{article}

\begin{document}

\title{Contribution of Extragalactic Infrared Sources
     to\\ CMB Foreground Anisotropy}

\author{Eric Gawiser and George F. Smoot}
\affil{Department of Physics and Lawrence Berkeley National Laboratory, 
   University of California, Berkeley, Berkeley, CA 94720}
\authoremail{gawiser@physics.berkeley.edu}

\begin{abstract}

We estimate the level of confusion to 
Cosmic Microwave Background (CMB)
anisotropy measurements
caused by extragalactic infrared sources.  CMB anisotropy observations at 
high resolution and high frequencies are especially sensitive to this 
foreground.
We use data from the COBE satellite to generate a Galactic emission spectrum
covering mm and sub-mm wavelengths.  Using this spectrum 
as a template, we predict the
microwave emission of the 5319 brightest infrared galaxies seen by IRAS.
We simulate skymaps of extragalactic infrared sources over the relevant 
range of frequencies (30-900 GHz) and instrument resolutions 
($10'-10^{\circ}$ FWHM).  
Analysis of the temperature anisotropy of these skymaps shows 
that a reasonable observational window is available for 
CMB anisotropy measurements.

\end{abstract}


\section{Introduction}

	The COBE detection of large-angular scale Cosmic Microwave Background
anisotropy (Smoot et al. 1992) has generated
interest in 
measuring CMB anisotropy on all angular scales with the goal
of determining cosmological parameters.  
Current anisotropy observations look at  sub-degree angular scales 
which correspond to observable structures
in the present universe.  Improved instrumentation
and the MAP (Microwave Anisotropy Probe) and Max Planck Surveyor
(formerly COBRAS/SAMBA) satellite missions focus attention on 
angular scales between
one-half and one-sixth of a degree.

Due to its large beam size, COBE was basically unaffected by 
extragalactic foreground sources (Banday et al. 1996, Kogut et al. 1994).
Because the antenna temperature 
contribution of a point source increases with the inverse 
of the solid angle of the beam,
observations at higher 
angular resolution 
are more sensitive to extragalactic foregrounds, including 
radio sources, the Sunyaev-Zel'dovich 
effect from galaxy clusters, and the infrared-bright galaxies examined
here.
   
	Previous work in this area (Toffolatti et al. 1994, 
Franceschini et al. 1989, Wang 1991)
 used galactic evolution models with specific 
assumptions about dust
temperatures to predict the level of extragalactic foreground.  
We choose instead a phenomenological approach using 
the infrared-bright galaxies 
detected by the Infrared Astronomical Satellite (IRAS) and the Galactic 
emission detected by the COBE (Cosmic Background Explorer) satellite. 
Section 3 compares our results with those from galaxy-evolution models.  

	The FIRAS (Far-Infrared Absolute Spectrophotometer) instrument of COBE
gives evidence for the existence of 
Cold ($<15$K) Dust in the Galactic plane (Reach et al. 1995).  
If the Milky Way has Cold Dust, then it is likely present in 
other dusty spirals, which comprise the majority of bright 
extragalactic infrared sources.  Some observations (Chini et al. 1995, 
Block et al. 1994, Devereux \& Young 1992) 
indicate the presence of Cold Dust in other
galaxies.  Neither galactic evolution models nor pre-FIRAS observations (see
Eales et al. 1989) were able to set tight constraints on emission from Cold 
Dust, but the FIRAS observations do.  
Emission from dust close in temperature to 
the 2.73 K background radiation
is difficult to separate from real CMB anisotropies. 
If Cold Dust is typically accompanied by Warm Dust in spiral galaxies, 
we can use
the FIRAS information about the total dust emission of the Galaxy to overcome
this spectral similarity.

\section{Extragalactic Infrared Sources}	

	The far-infrared discrete sources 
detected by IRAS are typically inactive spiral 
galaxies, although some are quasars, starburst galaxies, and Seyfert galaxies.  
The IRAS 1.2 Jy 
catalog (Fisher et al. 1995) 
provides flux measurements of 5319 galaxies at 12, 25, 60, and 100 $\mu$m, 
where interstellar dust emission is dominant.  
We compared the locations of these galaxies 
with those of a thousand of
the brightest radio sources, and only 7 possible coincidences resulted.  
This lack of coincidence shows that radio-loud galaxies can
be treated separately.  
The IRAS sources are roughly isotropic in distribution, except for
a clear pattern of the Supergalactic Plane.               
To reduce the
possibility of residual galactic contamination, we use a skymap in our analysis
that covers galactic latitudes $|b| > 30^{\circ}$. 
This map contains contributions from 2979 galaxies for a $0\fdg5$ beam.  

The nature of dust in spiral galaxies is still an open question.
It seems likely 
that there is dust at widely 
varying temperatures and possibly with different emissivities (Rowan-Robinson 
1992, Franceschini \& Andreani 1995).    
Attempts to fit observational data have yielded a variety of 
results; it is unclear if far-infrared luminous dust is well described by 
a one-component or a two-component model, and the emissivity power-law 
index is 
only known to be between 1 and 2.  We avoid specifying 
the nature of this dust by using the observed Galactic far-infrared emission
spectrum as a template for IRAS galaxies.
To check the accuracy of this template, we fit 
a two-component dust model to IRAS galaxies and
to the integrated 12, 25, 60, and 100 $\mu$m 
fluxes of the Milky Way measured by the DIRBE (Diffuse Infrared Background
Experiment) instrument of COBE.  This produces similar 
results for the Warm (15-40 K) Dust component to which IRAS and 
DIRBE are most 
sensitive; for an emissivity power-law index of 1.5, DIRBE gives
a Warm Dust temperature of 28K for the Milky Way, while the 425 
IRAS galaxies with highest-quality flux measurements are 
collectively fit to a Warm Dust temperature of 33K.  This Warm Dust 
accounts for the majority of the far-infrared emission of spiral galaxies.

There is, however, observational evidence that the far-infrared emission
of inactive 
spirals is dominated by dust slightly colder than 20K 
(Neininger \& Guelin 1996, Chini \& Krugel 1993).  Fitting the FIRAS 
spectrum of the Milky Way also leads to a Warm Dust temperature close to
20K.  These fits appear to conflict with the temperatures found above using
IRAS and DIRBE fluxes at $\lambda \leq 100 \mu$m.  
Using 
60, 100, 140, and 240 $\mu$m DIRBE fluxes, however, indicates a
Warm Dust temperature for the Galaxy of 24K.  This shows that temperature
fits to data on one side of 
the peak of the assumed blackbody spectrum can be 
inaccurate.  
Figure 1 shows that
the spectra of the Milky Way found by DIRBE and FIRAS
are indeed
compatible.  It may be an oversimplification to represent the 
Warm Dust in a galaxy by a single temperature.

We recognize that 
not all IRAS galaxies have exactly the same far-IR spectrum as the Milky Way.  
Active galaxies are warmer, 
with an average Warm Dust temperature of 33K (for emissivity index 2, Chini
et al. 1995).  However, the
cirrus emission which dominates Galactic dust is consistent
with the emission from the majority of inactive spirals (Andreani \& 
Franceschini 1996, Pearson \& Rowan-Robinson 1996).   
Some observations indicate that our Galaxy is slightly warmer than 
the average inactive spiral (Chini et al. 1995).  None of these observations
includes enough frequencies to provide a 
template microwave emission spectrum, and
their results range by a factor of 3 depending on the choice of beam
corrections (Franceschini \& Andreani 1995).    
The Milky Way is a 
good middle-of-the-road choice for a microwave template spectrum;
the 
DIRBE and IRAS dust 
temperature fits given above 
agree rather well.

After removing Galactic emission lines (as in Reach et al. 1995), 
we fit a dust model to the 
FIRAS dust spectrum.  The CO 1-0 emission line at 115 GHz is not clearly
detected by FIRAS but could be responsible for increased
emission at that frequency. 
It is possible to vary the parameters of the 
dust model significantly and still have an acceptable fit, so we refrain 
from giving any physical importance to the parameters of the fit.  
We add synchrotron and free-free components with microwave-range
spectral indices of 
$-1.0$ and $-0.15$, 
respectively, so that these sources of microwave emission match COBE DMR
(Differential Microwave Radiometer)
observations below 100 GHz (Kogut et al. 1996, Reach et al. 1995, 
Bennett et al. 1992).  
Free-free emission is stronger than dust beyond the low-frequency end
of the FIRAS spectrum.     

We combine data from DIRBE, FIRAS, and DMR to form the  
broad Galactic spectrum shown in Figure 1.  Each IRAS 1.2 Jy source is fit to 
the DIRBE end of the spectrum and extrapolated to the desired frequency using 
this template. 
In fitting each IRAS galaxy to the DIRBE fluxes of the Milky Way,
we give more weight to the 60 and 100 $\mu$m fluxes, which are most 
sensitive to Warm Dust, than to the 12 and 25 $\mu$m fluxes, 
which are also sensitive to Hot (100-300 K) Dust.  
The 1.2 Jy catalog gives redshifts for these galaxies.  
Most have $z < 0.05$ and all have $z < 0.3$. 
We take these redshifts into account while fitting 
and extrapolating.

It would be advantageous to 
fit each type of galaxy
to a specialized far-IR to microwave spectrum, but no other 
trustworthy template spectrum is currently available, 
so we use the Galactic far-infrared emission spectrum for all sources.
The Galactic spectrum agrees well with observed correlations between radio
and IR fluxes of IRAS galaxies 
(Condon \& Broderick 1991, Crawford et al. 1995).
Our template spectrum is consistent 
with detections and 
upper limits for bright infrared galaxies
from DIRBE (Odenwald, Newmark \& Smoot 1995).  
This is helpful because DIRBE used 140 and 240 $\mu$m
channels, which IRAS lacks, allowing it to probe much cooler
dust temperatures than IRAS.
DIRBE rules out the 
possibility of extremely bright sources occurring in the 2\% of the high 
Galactic latitude sky not surveyed by IRAS and sees no evidence for 
sources whose emission comes predominantly from Cold Dust.

\section{Results}

We use the Galactic far-infrared emission spectrum to predict the 
microwave flux of each IRAS galaxy in Jy ($1$ Jy $= 10^{-26} W/m^2/$Hz).  
To convert from flux $S$ to antenna temperature $T_A$, we use

\begin{equation}
 T_A = S \frac{\lambda^2}{2 k_B \Omega}\; \;,  
\end{equation}

\noindent 
where $k_B$ is Boltzmann's constant, $\lambda$ is the wavelength, and 
$\Omega$
is the effective beam size of the observing instrument.  Antenna temperature
is related to thermodynamic temperature by

\begin{equation}
 T_A = \frac {x}{e^x - 1} \; \; T,  
\end{equation}

\noindent
defining $x \equiv h \nu / k T$.  Small fluctuations in antenna 
temperature can be converted to effective thermodynamic 
temperature fluctuations using

\begin{equation}
 \frac{dT_A}{dT} = \frac { x^2 e^x} {(e^x - 1)^2} \; \; .  
\end{equation}

Analysis of source counts indicates that the 1.2 Jy sample is complete down to 
an extrapolated flux of 3 mJy at 100 GHz.  We divide 
the sources logarithmically 
into groups of similar flux and 
find a gradual 
decrease in anisotropy as flux decreases, indicating that dimmer sources
will not generate significant anisotropy.  Toffolatti et al. (1995) 
found a negligible contribution from non-Poissonian fluctuations.  
Poissonian fluctuations should
be dominated by those sources prevalent enough to have roughly one source
per pixel.  
For an instrument with a resolution of $10'$ to have one source
per beam, we must look at sources with $ z \simeq 0.24$.  
Assuming $(1+z)^3 $ luminosity evolution and including 
k-correction  
(see Pearson \& Rowan-Robinson 1996, Beichman \& Helou 1991),
these sources will generate a 
temperature anisotropy only 2\% of that caused by IRAS 1.2 Jy galaxies.
High redshift galaxies 
should produce a significant isotropic cosmic infrared background
(Hauser 1995)
but should be too distant to
produce significant foreground anisotropy.  
We therefore expect the anisotropy generated 
by sources too dim to make the 1.2 Jy catalog to be a small part of the total
anisotropy; the brightest sources are generating most of the fluctuations.

	To simulate observations, we 
convolve all sources on pixelized skymaps (2X oversampled) of
resolution varying from $10'$ to $10^{\circ}$. 
The resulting maps, covering a range of frequencies from 30 to 900 GHz, are
analyzed to 
determine the expected contribution of IRAS galaxies to foreground 
confusion of CMB temperature anisotropy.
The information contained in these skymaps can be used 
to choose regions of the sky in which 
to observe (Smoot et al. 1995).  
The contour plot in 
Figure 2 shows the 
rms thermodynamic temperature anisotropy
produced by extragalactic infrared sources 
over the full range of frequencies and instrument resolution.  
The minimum value of $\frac{\Delta T}{T}$
is $1.3\times 10^{-8}$ at large FWHM and medium frequency 
and the maximum value is $0.092$ 
at small FWHM and high frequency.  For frequency in GHz and FWHM in degrees, 
our results for temperature
anisotropy are fit to within 10\% by

\begin{equation}
\log_{10}\frac {\Delta T}{T} = 2.0 (\log_{10}\nu)^3 - 8.6 (\log_{10}\nu)^2
+ 10.3 \log_{10}\nu - 0.98 \log_{10} (FWHM) - 9.2 \; \; \; 
\end{equation}

\noindent The inverse linear relationship between anisotropy and
FWHM results from the combined effects of beam convolving and map pixelization.
Anisotropy from extragalactic infrared sources dominates 
expected CMB anisotropy at frequencies above 500 GHz.  This makes effective
foreground discrimination possible for instruments with a frequency range 
sufficiently wide to detect the extragalactic infrared 
foreground directly.

Figure 3 shows a summary of our results for several benchmark instrument
resolutions.  The dashed lines represent 
the results of subtracting pixels where
the fluctuations from extragalactic infrared sources are five times
times greater than the quadrature sum of the rms CMB anisotropy and the 
expected
instrument noise for the Max Planck Surveyor at that frequency
(Tegmark \& Efstathiou 1995).
These $5 \sigma$ pixels can be
assumed to contain bright point sources. 
Our results agree closely with those of 
Toffolatti et al. (1995) for their
model of moderate cosmological evolution of all galaxies.  
Our predictions for anisotropy are lower by about a factor of three 
than those
of Franceschini et al. (1989), 
who assume strong evolution of the 
brightest IR sources and include early galaxies with heavy starburst activity. 
Wang (1991) ignores the possibility of cold dust and uses galaxy evolution 
models to predict anisotropy levels somewhat lower than those found 
with our phenomenological approach.  

The $5 \sigma$ subtraction has a significant  
effect for small FWHM at frequencies below 500 GHz.   
The maximum effect is to subtract 0.002\% of the pixels, 
leading to a factor of 5 reduction in foreground 
temperature anisotropy.  This is
further 
evidence that temperature anisotropy from extragalactic infrared sources 
is dominated by the brightest sources.
  The bright sources are a mixture of Local Group galaxies and 
more distant infrared-luminous galaxies such as starburst galaxies.  Optimal
subtraction of the extragalactic infrared foreground requires the contribution
from each bright source to be predicted accurately.

\section{Discussion}

Our usage of the Galactic far-infrared emission spectrum as a template
causes systematic errors on a galaxy-by-galaxy basis.  
It is easy to place constraints on our results; if all
galaxies had only 33K dust as is typical for active galaxies, the resulting 
anisotropy would be a factor of 100 lower.  This is highly unlikely, because we
know that most IRAS galaxies are inactive spirals, and galaxies
with colder dust will dominate the anisotropy at mm-wavelengths because 
of the selection effect favoring sources with flatter spectra.  A robust upper
limit on microwave anisotropy from infrared galaxies can be set by assuming
that our IRAS 1.2 Jy galaxies cause the full cosmological far-infrared
background (for which Puget et al. 1996 claims a detection and 
Mather et al. 1994 gives an upper limit).  In this case we have underestimated
the anisotropy by a factor of 100, but no predictions of the IR background
expect these nearby galaxies to produce more than a few percent of it.  
A more realistic check on our results comes from Andreani \& Franceschini
(1995), who measured a complete sample of IRAS galaxies at 1300 $\mu$m 
(240 GHz).  Their average flux ratio of 1300 $\mu$m over 100 $\mu$m is half
that of the Galaxy, but one of their beam correction methods brings their ratio
into agreement with the Milky Way.  They find that the 60 $\mu$m 
emission of spiral galaxies receives enough contribution from a starburst
dust component mostly absent in the Galaxy that including 60 $\mu$m fluxes
in our fits may have caused a factor of 2 overestimate.  
Combined, these corrections
give us a possible systematic overestimate of anisotropy by a factor of 4.  If
typical IR-bright galaxies have dust colder than the Milky Way, our
results could instead be an underestimate by a factor of a few, but this
appears less likely. 
Our total systematic error is probably less than a factor of 3, which is 
consistent with our good agreement with previous results discussed in Section
3.
  
	The recently obtained spectral knowledge of our Galaxy has enabled
us to take into account the possible 
presence of Cold Dust.  
Our predicted level of temperature anisotropy makes the extragalactic
infrared foreground dominant over the Galactic foregrounds
of dust, free-free, and synchrotron for angular resolutions 
near $10'$ and frequencies above 100 GHz.  
Below 100 GHz, radio sources are
expected to be the dominant extragalactic
foreground.  
The extragalactic infrared foreground will not be significant in comparison to
CMB anisotropies around 100 GHz but will be dominant above 500 GHz.  
  Despite the likely presence of Cold Dust in infrared-bright galaxies,
our results leave a window at intermediate frequencies
for the measurement of 
CMB anisotropies without significant confusion
from extragalactic infrared sources.

\section{Acknowledgments}

We thank Michael Strauss and Dave Schlegel for their 
help with the IRAS 1.2 Jy catalog.
The DIRBE integrated 
fluxes of the Milky Way were graciously provided by Ned Wright and Janet 
Weiland.  We also thank Bill Reach for 
supplying the FIRAS Galactic Dust spectrum,  
Marc Davis, Giovanni DeAmici, and Laura Cayon for helpful conversations, 
and Gianfranco DeZotti, Joe Silk, Ted Bunn, and Evan Goer 
for reviewing drafts of this paper. 
E.G. acknowledges the support of an NSF Graduate Fellowship.  
This work was supported in part at LBNL through DOE Contract No.
DE-AC03-76SF00098.

\clearpage

\begin{figure}
\figurenum{1}
\caption{ 
The FIRAS Galactic Dust spectrum, including emission lines, is 
shown by $+$ symbols.  The smooth curve is a fit to this spectrum 
based upon a two-component dust model with synchrotron and free-free emission 
included using DMR results.  With a $\nu^2$ emissivity law
assumed, the fit is Warm Dust at $19.4$K and Cold Dust at $4.3$K with an 
optical depth 12.1 times that of the Warm Dust.  
The FIRAS error bars are not shown because they are
extremely small on this scale.  
The open circles are 
DIRBE integrated Galactic fluxes at 12, 25, 60, 100, 140, and 240 $\mu$m,
normalized to the FIRAS measurements.
}
\end{figure}

\begin{figure}
\figurenum{2}
\caption{
A log-log-log contour plot 
of equivalent thermodynamic temperature anisotropy 
due to extragalactic infrared sources as a function of frequency in GHz 
and angular resolution (FWHM) in degrees.  
The temperature anisotropy shown is $\log_{10} \frac{\Delta T}{T}$
where $\Delta T$ is the root mean square equivalent thermodynamic
temperature generated
by extragalactic infrared sources in Kelvin and $T$ is the 
temperature (2.73K) of the CMB.  The increase in anisotropy at low frequencies
occurs because synchrotron and free-free emission are included in our 
template spectrum. 
}
\end{figure}
              
\begin{figure}
\figurenum{3}
\caption{
Log-log plot of $\frac{\Delta T}{T}$
versus frequency for instrument resolutions of
$10'$, $30'$, $1^{\circ}$, and $10^{\circ}$,
showing window where foreground confusion should be
less than $10^{-6}$.  
Solid lines are for no pixel subtraction.  The 
dotted, dashed, and long-dashed lines show the results 
with pixels at a level of $5\sigma$ removed for resolutions of 
$10'$,$30'$, and $1^{\circ}$, respectively.  This $5\sigma$ subtraction 
makes no difference at any frequency for $10^{\circ}$.  
}
\end{figure}

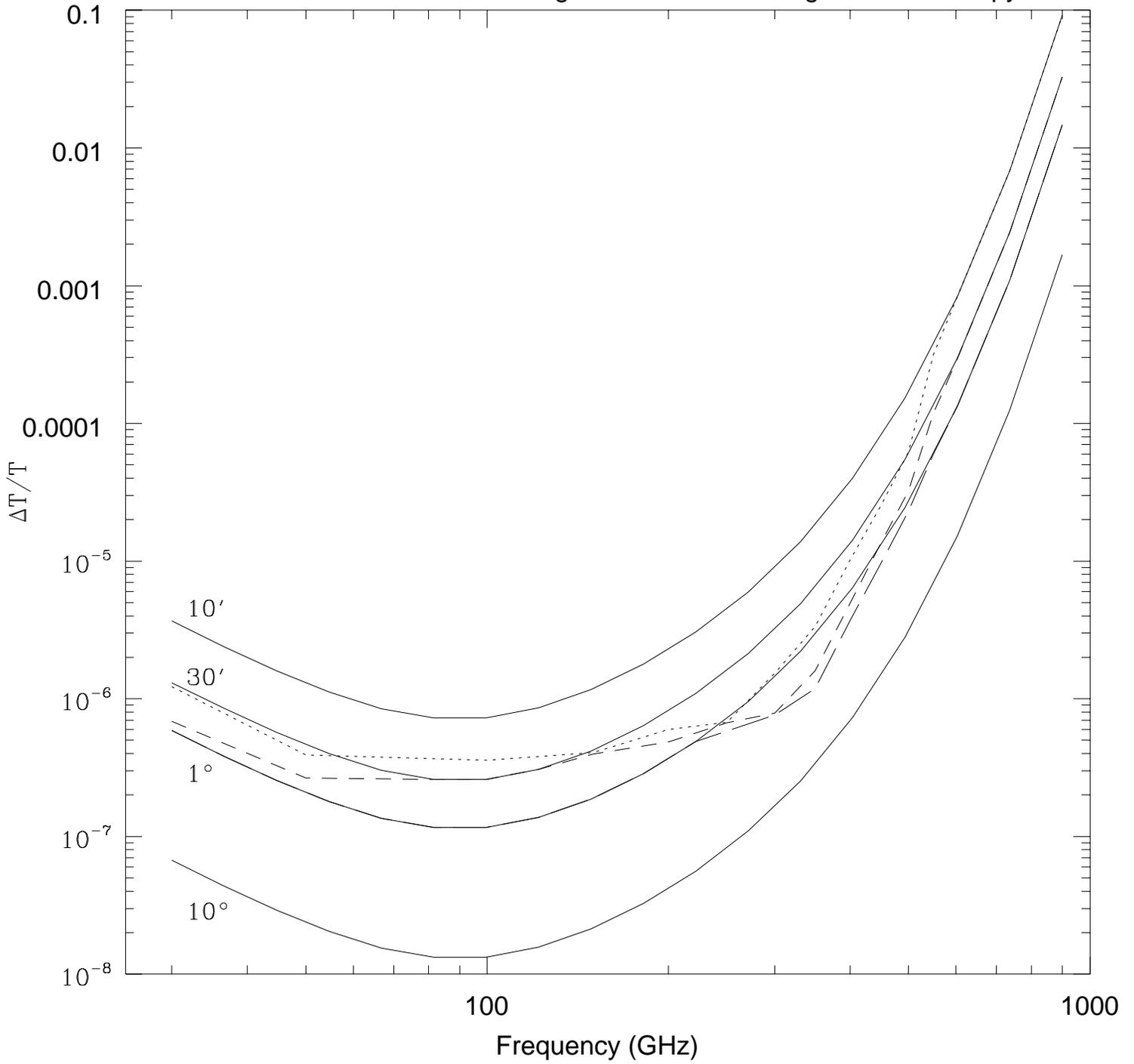
\end{document}